\documentclass[manuscript]{aastex631}
\shorttitle{YMCA-1}
\shortauthors{Gatto et al.}
\graphicspath{{./}{figures/}}

\begin{document}

\title{YMCA-1: a new remote star cluster of the Milky Way?
\footnote{This work is based on INAF-VST guaranteed observing time under ESO program: 0104.D-0427(A)}}

\correspondingauthor{Massimiliano Gatto}
\email{massimiliano.gatto@inaf.it}

\author[0000-0003-4636-6457]{M. Gatto}
\affiliation{INAF-Osservatorio Astronomico di Capodimonte, Via Moiariello 16, 80131, Naples, Italy}
\affiliation{Dept. of Physics, University of Naples Federico II, C.U. Monte Sant'Angelo, Via Cinthia, 80126, Naples, Italy}

\author{V. Ripepi}
\affiliation{INAF-Osservatorio Astronomico di Capodimonte, Via Moiariello 16, 80131, Naples, Italy}

\author[0000-0001-8200-810X]{M. Bellazzini}
\affiliation{INAF-Osservatorio di Astrofisica e Scienza dello Spazio, Via Gobetti 93/3, I-40129 Bologna, Italy}

\author{M. Tosi}
\affiliation{INAF-Osservatorio di Astrofisica e Scienza dello Spazio, Via Gobetti 93/3, I-40129 Bologna, Italy}

\author{C. Tortora}
\affiliation{INAF-Osservatorio Astronomico di Capodimonte, Via Moiariello 16, 80131, Naples, Italy}

\author{M. Cignoni}
\affiliation{INAF-Osservatorio Astronomico di Capodimonte, Via Moiariello 16, 80131, Naples, Italy}
\affiliation{Physics Departement, University of Pisa, Largo Bruno Pontecorvo, 3, I-56127 Pisa, Italy}
\affiliation{INFN, Largo B. Pontecorvo 3, 56127, Pisa, Italy}

\author[0000-0002-6427-7039]{M. Spavone}
\affiliation{INAF-Osservatorio Astronomico di Capodimonte, Via Moiariello 16, 80131, Naples, Italy}

\author{M. Dall'ora}
\affiliation{INAF-Osservatorio Astronomico di Capodimonte, Via Moiariello 16, 80131, Naples, Italy}

\author{G. Clementini}
\affiliation{INAF-Osservatorio di Astrofisica e Scienza dello Spazio, Via Gobetti 93/3, I-40129 Bologna, Italy}

\author{F. Cusano}
\affiliation{INAF-Osservatorio di Astrofisica e Scienza dello Spazio, Via Gobetti 93/3, I-40129 Bologna, Italy}

\author{G. Longo}
\affiliation{Dept. of Physics, University of Naples Federico II, C.U. Monte Sant'Angelo, Via Cinthia, 80126, Naples, Italy}

\author{I. Musella}
\affiliation{INAF-Osservatorio Astronomico di Capodimonte, Via Moiariello 16, 80131, Naples, Italy}

\author{M. Marconi}
\affiliation{INAF-Osservatorio Astronomico di Capodimonte, Via Moiariello 16, 80131, Naples, Italy}

\author{P. Schipani}
\affiliation{INAF-Osservatorio Astronomico di Capodimonte, Via Moiariello 16, 80131, Naples, Italy}







\begin{abstract}


We report the possible discovery of a new stellar system (YMCA-1), identified during a search for small scale overdensities in the photometric data of the YMCA survey. The object's projected position lies on the periphery of the Large Magellanic Cloud about $13\degr$ apart from its center. The most likely interpretation of its color-magnitude diagram, as well as of its integrated properties, is that YMCA-1 may be an old and remote star cluster of the Milky Way at a distance of 100 kpc from the Galactic center. If this scenario could be confirmed, then the cluster would be significantly fainter and more compact than most of the known star clusters residing in the extreme outskirts of the Galactic halo, but quite similar to Laevens~3. However, much deeper photometry is needed to firmly establish the actual nature of the cluster and the distance to the system.

\end{abstract}




\section*{}

YMCA ({\it Yes, Magellanic Clouds Again}, PI: V. Ripepi) is an optical survey carried out with the VLT survey telescope (VST), aimed at exploring the outskirts of the Large and the Small Magellanic Cloud (LMC and SMC, respectively), by collecting deep $g$ and $i$ images and photometry over an area of $\simeq 110$~deg$^2$ (Gatto et al., in preparation).
One of the aims of the YMCA survey is the search for unknown stellar systems, such as star clusters (SCs), in the periphery of the LMC/SMC. To this end, we adopted an automated algorithm based on kernel density estimation (KDE) that already proved successful in finding about 80 new LMC clusters in 23 sq. deg. of the surveyed area \citep[see][]{Gatto2020}.  

\begin{figure*}
    \includegraphics[scale=0.3]{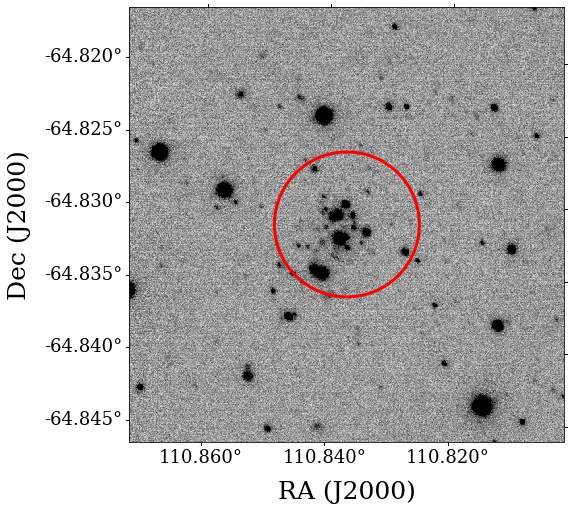}
    \includegraphics[scale=0.35]{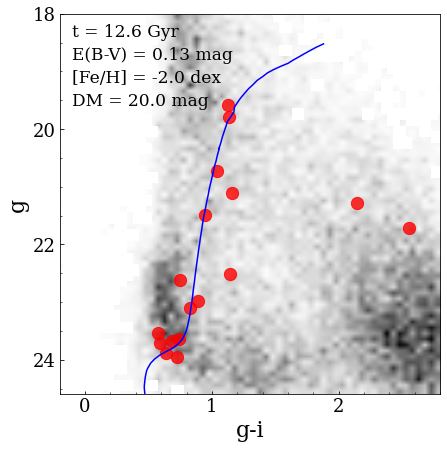}
    \includegraphics[scale=0.35]{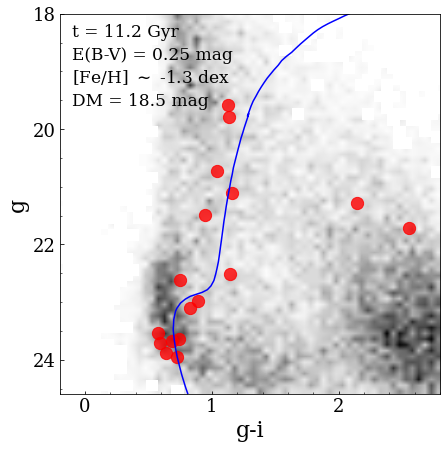}
    \caption{\emph{Left:} $g-$band sky image of $1.5\arcmin\times1.5\arcmin$ in size centred on YMCA-1. The seeing is 0.9$\arcsec$ FWHM. The red circle has a radius of 0.3$\arcmin$. \emph{Right:} CMD of a $1\degr \times 1\degr$ region around YMCA-1 (gray shaded) with stars within 0.3$\arcmin$ from the YMCA-1 centre, shown as red filled circles, and best fitting isochrones of the YMCA-1 CMD (blue solid lines), whose parameters are reported in the top left corner of the panels.}
    \label{fig:ymca_1}
\end{figure*}

Extending this analysis to newly reduced tiles of the YMCA sample, we discovered what appears as an uncatalogued stellar system placed about $13\degr$ to the East of the LMC centre. This over-density, that we have named YMCA-1, has a significance of 12.2 sigma over the local background and is easily visible as an agglomerate of stars centred at (RA, Dec) = (110.8369$\degr$,-64.8313$\degr$), in the image shown in the left panel of Fig.~\ref{fig:ymca_1}.
The corresponding colour-magnitude diagram (CMD) is displayed in the right panel of the same figure. Stars within 0.3$\arcmin$ from the system center are plotted as red filled circles, over the grey-scale Hess diagram of the population of the entire 1.0~deg$^2$ YMCA tile.
While the number of measured stars within the selected area is very small (15 stars), their distribution in the CMD is markedly different than that of the surrounding field. In particular, none of the candidate YMCA-1 members matches the most densely populated feature of the CMD attributable to the LMC, i.e. the Main Sequence (MS) Turn Off region in the range $22.2\la g\la 24.0$ mag, around $g-i=0.6$ mag, with the SGB at $g\simeq 22.3$ mag.

\begin{figure}
    \centering
    \includegraphics[scale=0.5]{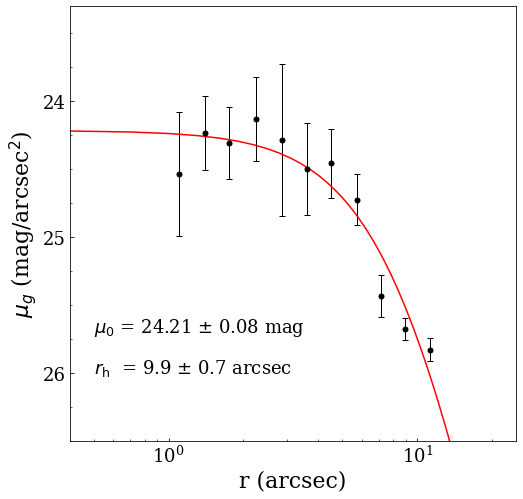}
    \caption{Surface brightness profile of YMCA-1. The red solid line represents the best-fit of a Plummer profile, whose parameters are displayed in the left lower corner of the figure.}
    \label{fig:plummer}
\end{figure}

\begin{figure*}
    \centering
    \includegraphics[scale=0.35]{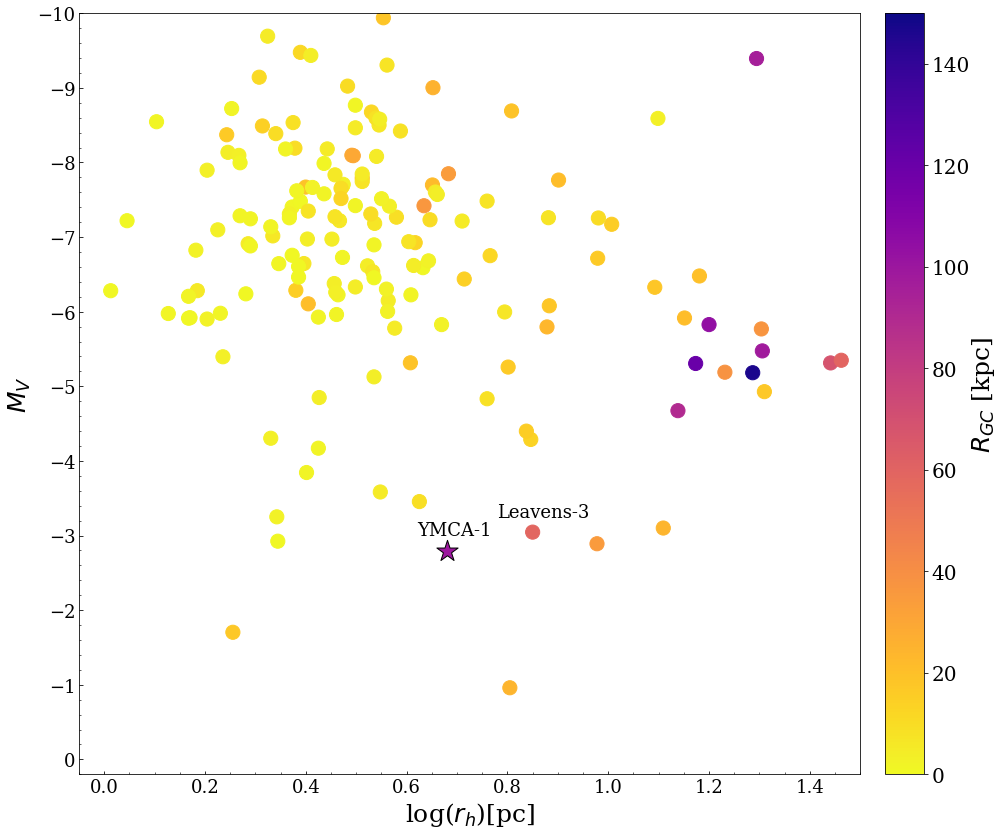}
    \includegraphics[scale=0.35]{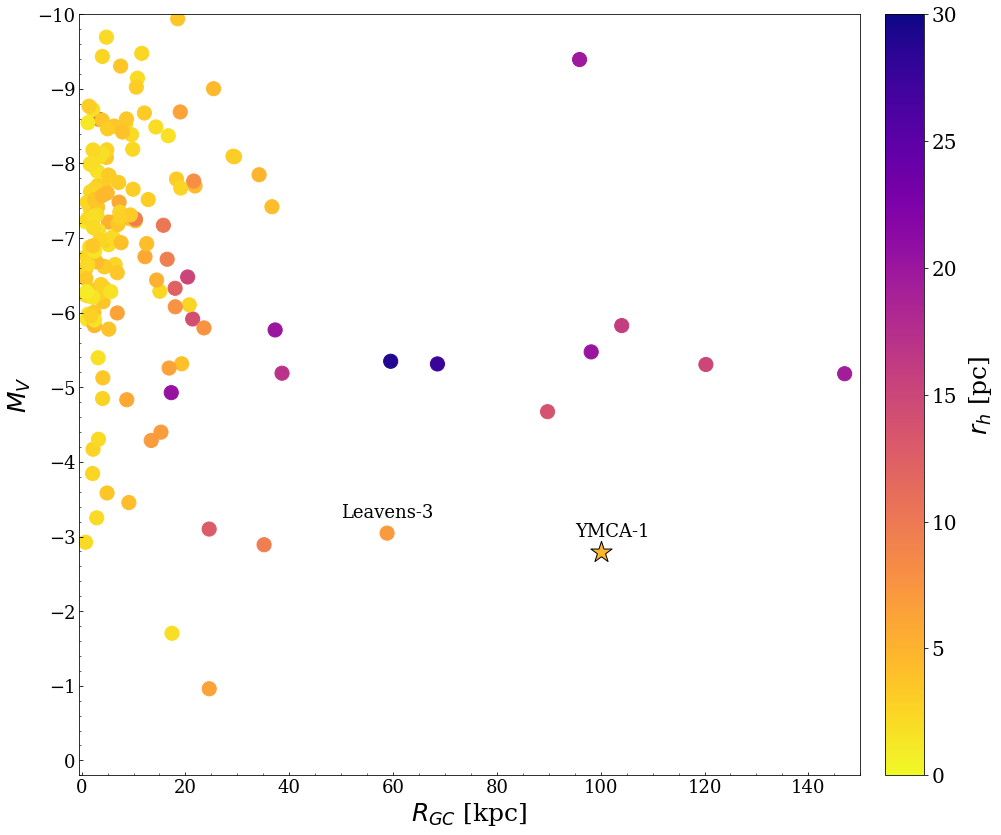}
    \caption{\emph{Top:} luminosity - half-light radius diagram, with points color-coded according their galactocentric distances.
    \emph{Bottom:} luminosity - galactocentric distances diagram, with points color-coded according their half-light radius. YMCA-1 is shown as a star in both panels.}
    \label{fig:mv_rh}
\end{figure*}

The two reddest putative YMCA-1 stars, at $g-i>2.0$ mag, are nearby interlopers according to their proper motions, as measured by the ESA/Gaia mission  \citep[Early Data Release 3 catalog, EDR3,][]{GaiaBrown2021}. Most of the remaining stars seem to be aligned along a steep Red Giant Branch (RGB), typical of old and metal poor stellar systems. The CMD of YMCa-1 can be fitted with an old (age=12.6~Gyr) and metal-poor ([Fe/H]=-2.0 dex) isochrone of the PARSEC set \citep[][]{Bressan2012}, shifted to the distance of $\simeq 105$~kpc and corrected for $E(B-V)=0.13$ (from the \citealt{SFD98} maps, recalibrated after \citealt{Schlafly2011}). In this solution, the distance estimate is mainly anchored to the interpretation of the faintest stars at $g\simeq 23.8$ mag as belonging to the system Sub Giant Branch (SGB). Adopting this set of parameters, we estimate $M_g\simeq -2.3$ mag (corresponding to $M_V\simeq -2.8$ mag) and $r_h\simeq 4.8$~pc, from a surface brightness profile obtained by integrated aperture photometry on concentric annuli, and fitted with a \citet[][]{Plummer1911} model, as displayed in Fig.~\ref{fig:plummer}.
The integrated color of YMCA-1 was derived through circular aperture photometry on both the $g$ and $i$ images, adopting a radius equal to 0.3$\arcmin$, obtaining $g - i = 1.1$ mag. The luminosity and color we estimated for YMCA-1 were used to estimate the total mass by means of the mass/luminosity - integrated color relationship available in \citet{Roediger2015}, finding a value of $\log M/M_{\odot} = 3.44$~dex. As a sanity check for the mass estimate, we used a synthetic CMD, with appropriate age and metallicity, built from the PARSEC isochrones \citep[][]{Bressan2012}, to derive the total mass of a stellar system with a similar number of RGB stars ($6 \leq N \leq 9$) as observed in YMCA-1. The estimated mass after 1000 random extractions is $\log M/M_{\odot} = 3.28 \pm 0.15$~dex, consistent with the value obtained through integrated properties.
In this scenario, YMCA-1 would thus be a remote Milky Way star cluster, with unusual properties compared to other star clusters at similar galactocentric distances. 
The uncertainty is admittedly very high, but alternative interpretations provide significantly worse fits to the observed CMD. For instance, assuming the system to be at the same distance of the LMC (D = 50 kpc), and letting the isochrone best fitting YMCA-1 fit also the LMC CMD would require assuming a reddening value much larger than that provided by the \citet{SFD98} reddening maps, i.e. E(B-V)=0.25 mag, as well as an isochrone with larger metallicity ([Fe/H]$\sim-1.3$, see right panel of Fig.~\ref{fig:ymca_1}).
The luminosity - half-light radius diagram is shown in the top panel of Fig.~\ref{fig:mv_rh}, while the luminosity - galactocentric radius is shown in the bottom panel of the same figure. YMCA-1 is plotted along with 161 Galactic globular clusters taken from the \citet{Baum2018} catalog. We retrieved reddening values from \citet[][ 2010 version]{Harris-1996}, with some exceptions as listed in Tab.~\ref{tab:redd}. YMCA-1 could be one of the faintest star clusters ever discovered hitherto and definitely the most compact beyond 50 kpc.
The \citet{Baum2018} catalog lists only nine globular clusters (GCs) beyond a galactocentric distance of $R_{GC}$=50~kpc, three of them discovered in the last decade thanks to deep multi-band photometric surveys. With the remarkable exception of NGC~2419 ($M_V=-9.4$ mag), they are all relatively sparse clusters. Still, seven of them are significantly more luminous ($-4.7\le M_V\le-5.8$ mag) and more extended (13.7~pc$\le r_h\le$29.0~pc) than YMCA-1, in the hypothesis that it is a distant star cluster of the MW. On the other hand, Laevens~3 \citep{Longe2019}, at $R_{GC}=58.8$~kpc, has structural properties quite similar to YMCA-1, i.e. $M_V = -3.1$ mag and $r_h=7.1$~pc.
Such relatively compact but very faint star clusters are very difficult to spot at such large distances from us, suggesting that several more may be still lurking in the remotest region of the Galactic halo
\citep[see also][]{Webb2021}.
However, before any firm conclusion can be drawn, follow-up with deeper photometry is required to  
unambiguously detect the MS of YMCA-1, to confirm its nature and reliably estimate its distance.
\begin{table}[]
    \centering
    \begin{tabular}{c|c}
    \hline
    Cluster & Reference\\
    \hline
    BH140 & \citet{SFD98}\\
    Crater & \citet{Weisz-2016}\\
    FSR1716 & \citet{Minniti-2017}\\
    FSR1758 & \citet{SFD98}\\
    Leavens3 & \citet{Laevens-2015}\\
    RLGC1 & \citet{Ryu-2018}\\
    RLGC2 & \citet{Ryu-2018}\\
    Sagittarius II & \citet{Laevens-2015}\\
    VVV-CL001 & \citet{Fernandez-Trincado2021}\\
    \hline
    \end{tabular}
    \caption{List of the GCs whose extinction values were not present in \citet{Harris-1996}, and their references. Reddening values taken from \citet{SFD98} have been re-calibrated by \citet{Schlafly2011}.}
    \label{tab:redd}
\end{table}

\facilities{VST}


\software{sklearn \citep[][]{scikit-learn},   
photutils \citep[][]{bradley2019}, astropy \citep[][]{astropyI,astropyII}, matplotlib \citep[][]{matplotlib}, AstroWise \citep{AstroWise} }




\bibliography{mybibliography_sbp}{}
\bibliographystyle{aasjournal}



\end{document}